\documentclass[authoryear,12pt,letter,3p]{jowarticle}
\usepackage{graphicx}
\usepackage{amsmath}
\usepackage{amssymb}
\usepackage{setspace}
\usepackage{appendix}
\usepackage[all]{xy}
\makeatletter
\renewcommand\section{\@startsection{section}{1}{\z@}{-3.25ex plus -1ex minus -.2ex}{1.5ex plus .2ex}{\normalsize\bf}}
\renewcommand\subsection{\@startsection{subsection}{2}{\z@}{-3.25ex plus -1ex minus -.2ex}{1.5ex plus .2ex}{\normalsize\bf}}
\renewcommand\subsubsection{\@startsection{subsubsection}{3}{\z@}{-3.25ex plus -1ex minus -.2ex}{1.5ex plus .2ex}{\normalsize\bf}}
\makeatother

\begin{document}
\begin{frontmatter}
\title{Equivalence and Duality in Electromagnetism}

\author{James Owen Weatherall}\ead{weatherj@uci.edu}
\address{Department of Logic and Philosophy of Science \\ University of California, Irvine}

\date{ }

\begin{abstract}
In this paper I bring the recent philosophical literature on theoretical equivalence to bear on dualities in physics.  Focusing on electromagnetic duality, which is a simple example of S-duality in string theory, I will show that the duality fits naturally into at least one framework for assessing equivalence---that of categorical equivalence---but that it fails to meet a necessary condition for equivalence on that account.  The reason is that the duality does not preserve ``empirical content'' in the required sense; instead, it takes models to models with ``dual'' empirical content.  I conclude by discussing how one might react to this.
\end{abstract}
\end{frontmatter}
\doublespacing
\section{Introduction}

The past decade has seen a surge of interest in the (in)equivalence of theories among philosophers of science.\footnote{For a review of this recent literature, see \citet{WeatherallPart1,WeatherallPart2}.}  During the same period, philosophers of physics have begun to study a theoretical relationship, widely discussed by physicists, known as ``duality''.\footnote{For a representative sample of this literature, see \citet{RicklesOld,Rickles}, \citet{Matsubara}, \citet{Dieks+etal}, \citet{Read}, \citet{Dawid}, \citet{Huggett}, \citet{deHaroDEG,deHaro}, and \citet{deHaro+Butterfield}.  It is perhaps helpful to remark that de Haro, in various places, uses the expression ``theoretical equivalence'', but his meaning is different from how it is used in the theoretical equivalence literature.  (His expression ``physical equivalence'' is closer to what others have meant by ``theoretical equivalence''.)}   Physicists working in this subject often characterize duality as a kind of equivalence between distinct descriptions of the world---a characterization that philosophers of physics have tended to endorse as well.  And yet, the (philosophical) literature on duality and the (philosophical) literature on equivalence have proceeded largely in parallel, with very little contact.

My goal here is to bring these two literatures together, and to make a simple (though perhaps controversial) claim: Dualities are not (generally) theoretical equivalences.\footnote{\citet{Butterfield} has also very recently brought these literatures together---and indeed, his position is very close to mine, insofar as he argues that dualities are \emph{not} (necessarily) theoretical equivalences because (for instance) they can make contradictory claims about the same subject matter.  (I also note that Butterfield's view, here, is closely connected to ideas defended for several years by de Haro and collaborators \citep{Dieks+etal,deHaroDEG,deHaro+Butterfield}---though those authors do not make the connection to the theoretical equivalence literature.)  I think this is entirely correct.  Still, I have an important quibble with Butterfield.  He takes dualities to be formal equivalences, and then suggests that philosophers of science working on theoretical equivalence have taken such formal equivalences to be sufficient for theoretical equivalence.  He concludes that dualities \emph{would} be equivalences by the lights of the criteria philosophers have discussed.  He goes on to argue---along lines similar to \citet{Sklar}, \citet{Coffey}, and \citet{Nguyen}---that dualities thus show that formal criteria of equivalence are unacceptable because they do not capture the role of interpretation in assessing whether theories may be inequivalent.  To the contrary, my position is that criteria such as categorical equivalence, as proposed for instance by \citet{WeatherallTE} and applied by \citet{Rosenstock+etal}, explicitly includes a requirement that the mapping realizing a putative equivalence \emph{preserve empirical content} in a suitable way, and that it is because dualities may fail to satisfy this requirement that they are not equivalences.  In other words, insofar Butterfield's goal was to criticize the theoretical literature, I think he has missed the mark.  On the other hand, I think his substantive remarks on dualities are entirely convincing.}  The reason is that dualities do not preserve \emph{empirical} content---at least, not in the sense that has usually been required in the theoretical equivalence literature---and philosophers have generally taken empirical equivalence to be a necessary, though not sufficient, condition for theoretical equivalence of any sort.

This claim might seem surprising, at least if one is familiar with the literature in philosophy of physics on dualities, where the claim that dualities \emph{do} preserve empirical content (and, often, a whole lot more) is regularly made.  For instance, in an early philosophical foray into string theory, Dean Rickles writes, ``Roughly, two theories are dual whenever they determine the same physics: same correlation functions, same physical spectra, etc. ... [D]ualities map one theory onto another in a way that preserves all `physical' predictions'' \citep[p. 55]{RicklesOld}.  Similarly, and more recently, string theorist-turned-philosopher Richard Dawid writes, ``String dualities establish empirical equivalence between theories that often look entirely different with respect to their basic ontology and physical structure \citep[p. 21]{Dawid}.\footnote{For one more example, Baptiste Le Bihan and James Read \emph{define} a ``duality'' as an empirical equivalence: ``[A] duality is a mapping---that is, a systematic correspondence---between the spaces of solutions of two theories, such that models related by that map are empirically equivalent \citep[p. 2]{LeBihan+Read}.\nocite{Polchinski}} 

I will argue in what follows that all of these distinguished philosophers of physics---among many others---are wrong about whether dualities realize empirical equivalences.  But there is an important proviso: my argument will be that dualities do not preserve empirical content \emph{in a particular sense} that has been required in the theoretical equivalence literature.  Hence, my arguments, if sound, can be used to support two different positions: either that dual theories are not really equivalent; or that philosophers of science working on theoretical equivalence have gotten empirical equivalence wrong.  I will give reasons why I think the first of these is more attractive, but I take it that both positions would be of interest.

The remainder of the paper will be structured as follows.  I will begin by describing empirical equivalence as it has been understood in the theoretical equivalence literature.  I will then introduce a particular example of a duality---electromagnetic duality, which is a special case of what is known as \emph{S-duality} or \emph{Montonen-Olive duality}---to which a widely discussed criterion of theoretical equivalence can be readily applied.  As I will show, electromagnetic duality does not preserve empirical content.  I will conclude by discussing what one would need to do to recover empirical equivalence in this case, and explore why one might not be satisfied by such a move.

\section{Theoretical Equivalence}

Over the last half decade, philosophers of science have considered several potential criteria for theoretical equivalence.  For instance \citet{GlymourTETR,GlymourTE} proposed that \emph{definitional equivalence}, a notion of equivalence borrowed from first order logic, provides a natural criterion of equivalence of theories.  Two theories are definitionally equivalent if, roughly speaking, the non-logical vocabulary of each theory can be defined using the non-logical vocabulary of the other, in such a way that any true assertion of one theory can be translated into a true assertion of the other, and vice versa, in an invertible manner.  To avoid certain technical limitations of definitional equivalence, related to multi-sorted logic, \citet{Barrett+HalvorsonME} have proposed a generalization of definitional equivalence that they call \emph{Morita equivalence}.  \citet{Halvorson} and \citet{WeatherallTE}, meanwhile, have proposed \emph{categorical equivalence} (described below), a notion of equivalence borrowed from category theory, as a more attractive, more ``structural'' notion of equivalence.  And so on.

These criteria are generally named for mathematical relationships between (formal representations of) theories.  But these mathematical relationships are taken to be necessary, but not sufficient, conditions for equivalence.\footnote{To be fair, Glymour does not discuss this point in the earliest work in the literature; it was \citet{Sklar} who first emphasized it.  But in subsequent work, this condition is explicitly included in statements of the relevant criteria.}  For two physical theories to be theoretically equivalent, on any of the proposals just mentioned, one also requires that they be empirically equivalent, in a way that is compatible with the ``formal'' equivalence realized by the mathematical relationships.  More precisely, each of the mathematical criteria of equivalence under consideration involves mappings between formal structures that satisfy some stipulated conditions.  We also suppose that given a physical theory that may or may not be equivalent to another, there is some (fixed, prior) way of associating structures in the theories with the world.\footnote{One may think of this ``association'' as an intensional semantics, akin to what \citet[\S 3.1]{Butterfield} discusses (following Lewis); requiring such an association in advance seems to be what \citet{Dieks+etal} and \citet{deHaroDEG} characterize as adopting an ``external'' perspective.}  Empirical equivalence, in the present context, is the requirement that these mappings commute.

To illustrate what I mean, consider the case of categorical equivalence.\footnote{For background on category theory, see for instance \citep{Leinster}.}  To apply that criterion of equivalence to a pair of theories, we begin by defining categories $\mathcal{T}_1$ and $\mathcal{T}_2$.  The objects of each of these categories are models of each theory, respectively, and the arrows are appropriate ``structure preserving maps''.  These categories represent these theories.\footnote{How one goes from a ``theory in the wild'' to a category of models representing that theory is not always clear or unambiguous \citep{WeatherallWNCE,RosenstockThesis}.  This is sometimes a virtue, as in, for instance, \citep{BarrettCM2}, where different categories are used to characterize different interpretations of classical mechanics; but it can also highlight the fact that some work and prior understanding of the theories is necessary to even apply categorical equivalence---with the results often depending on what categories one chooses.}  In addition, one identifies a collection $W$ of ``stylized phenomena'' or ``predictions'' that one can associate with the theory, where each model of the theory will be associated with particular predictions corresponding to what one might observe (or what the world might be like) if the world were as represented by that model; we may take $W$ to be a category, trivially, by endowing each object with an identity arrow.\footnote{The details of how one does this are inessential to the point I am making.}  Finally, we fix functors $E_1:\mathcal{T}_1\rightarrow W$ and $E_2:\mathcal{T}_2\rightarrow W$, relating these models to the ``predictions'' of the theories.  Note that on this way of setting things up, it is already possible to associate the models of \emph{both} theories with predictions in a single category $W$; this is not as strong a condition as it appears, insofar as one could take $W$ to be the union of the ``stylized phenomena'' associated with each theory separately, in which case the functors $E_1$ and $E_2$ could still be defined, even though the theories made incomparable predictions.

To fix these ideas with an example, we might represent (vacuum) electromagnetism in Minkowski spacetime as follows.  We define a category $\mathcal{EM}$:
\begin{enumerate}
\item whose objects are Faraday tensors $F_{ab}$ on Minkowski spacetime satisfying the vacuum Maxwell equations; and
\item whose arrows are isometries $\chi$ of Minkowski spacetime that preserve $F_{ab}$.
\end{enumerate}
We may then take $W$ to consist of a specification of the empirical consequences of a particular Faraday tensor $F_{ab}$: its associated energy-momentum tensor, a collection of Lorentz force curves for test matter of different charge-to-mass ratios, etc.  Finally, we may define a ``functor'' $E:\mathcal{EM}\rightarrow W$ that takes Faraday tensors to the observable quantities associated with that tensor.

With this machinery in place, we may define categorical equivalence.  We will say that two theories are categorically equivalent if there exist functors $F_1:\mathcal{T}_1\rightarrow\mathcal{T}_2$ and $F_2:\mathcal{T}_2\rightarrow\mathcal{T}_1$ such that:
\begin{enumerate}
\item  $F_1$ and $F_2$ are essentially inverses, i.e., they realize an equivalence of categories; and
\item  $E_2\circ F_1 = E_1$ and $E_1\circ F_2 = E_2$.
\end{enumerate}
In other words, the categories representing the two theories are equivalent (as categories), and the functors realizing the equivalence commute with the maps taking models to their predictions.  Thus, if I begin with a model of the first theory and I ask what phenomena are associated with that model, and then I take the corresponding model of the second theory under the equivalence, it will be associated with the same predictions as the model with which I began.

\section{Electromagnetic Duality}

With this background from the theoretical equivalence literature in place, we now turn to dualities.  In fact, physicists discuss many different examples of dualities.  For instance, one widely discussed example that has played an important role in string theory since it was introduced more than two decades ago by \citet{Maldacena} is the AdS/CFT correspondence, also known as the gauge-gravity duality.  This duality relates a theory of gravity in an (asymptotically) anti-de Sitter space to a conformal field theory on the boundary of that space.  Another, even older, example is T-duality, which relates string theories defined on a manifold with radius $R$ to string theories defined on a manifold with radius $1/R$ \citep{tduality1,tduality2}.  Another, recently introduced, (family) of dualities are the particle-vortex dualities discovered by \citet{Karch+Tong} and \citet{Seiberg+etal}.  And so on.

My focus here is on yet another ``classic'' string duality, known as S-duality or Montonen-Olive duality \citep{Montonen+Olive,Seiberg+Witten,sduality,EMD}.  S-duality relates certain (supersymmetric) Yang-Mills theories whose coupling constants are large to other ones whose coupling constants are small.  This features makes the duality extremely useful, as strongly coupled theories are generally more difficult to work with than weakly coupled ones.  But it is also of interest for other, more foundational reasons.  For instance, it relates theories with gauge group $G$ to theories whose gauge group is the Langlands dual to $G$, which suggests a physical connection to the geometric Langlands program \citep{Frenkel}.

I will not consider S-duality in full generality.  Instead, I will focus on the simplest case of S-duality, known as electromagnetic duality.  The rest of what I say will apply to electromagnetic duality.\footnote{\citet[p. 209]{Dieks+etal} also briefly discuss electromagnetic duality, and adopt a perspective similar to the one defended here.}

At this point two concerns might arise.  First, one might worry that by focusing on just one example of a duality what I say will lack generality.  In fact, I am sympathetic to this worry, but I would go further: the word ``duality'' is used in many different ways in physics, and it is not at all clear that all ``dualities'' one encounters in the physics literature should (or can) be seen as instances of a single species.  Certainly one does not find a clear or precise definition of ``duality'' in the literature, or attempts to argue that such and such example satisfies such a definition.\footnote{Of course, philosophers of physics have attempted in recent years to fill this gap---see, especially, \citet{deHaro} and \citet{deHaro+Butterfield}.}  Still, I think the classic string dualities mentioned above---S-duality, T-duality, gauge-gravity duality, etc.---are meant to refer to importantly similar relationships, and I think, though I will not argue for this, that what I say about S-duality applies to these, at least, as well.\footnote{In \citep{tduality1}, for instance, the authors characterize the duality as relating each observable quantity in one theory to an observable quantity in the other---without, it seems, requiring that the correspondence preserve interpretation in any meaningful sense.  \citet{Butterfield}, likewise, takes this moral to apply to many dualities.}  But strictly speaking, whether this is true does not affect my argument, as my goal is to establish an existence claim: that is, there exist dualities that are not theoretical equivalences in the sense philosophers of science have considered.

The second concern is similar, but stronger: one might think that, by focusing just on \emph{electromagnetic} duality, I have moved away from the characteristic features of dualities.  I am not considering quantum field theories, much less string theory; I am not making use of any of the machinery usually used to show that a duality exists; etc.  To this I respond: true, but in these regards electromagnetic duality offers many advantages, especially regarding questions about empirical significance, because it is much clearer how one should understand classical electromagnetism to represent the world than string theory.  More, electromagnetic duality is widely cited as a (motivating) instance of S-duality \citep[cf. ][]{sduality,EMD}, and so one should expect the relevant features to be similar.  And, once again, even if this turns out to be false, my point is to make an existence claim.  To rebut it, one needs to either refute my argument concerning electromagnetic duality or else deny that electromagnetic duality is, indeed, a duality.  And this latter move, I suspect, would be difficult to make without also denying that electromagnetic duality is an instance of S-duality.

We now move on to define electromagnetic duality.  For simplicity, we work on Minkowski spacetime, $(M,\eta_{ab})$.  Mathematically, electromagnetic duality is a mapping from 2-forms to 2-forms given by:
\[
F_{ab}\mapsto \star F_{ab}
\]
where $\star$ is the Hodge star operation.  Now observe that this mapping preserves Maxwell's equations, in the sense that if $\nabla_a F^{ab}=\mathbf{0}$ and $\nabla_{[a}F_{bc]}=\mathbf{0}$, then $\nabla_a(\star F^{ab})=\mathbf{0}$ and $\nabla_{[a}\star F_{bc]}=\mathbf{0}$.\footnote{Something similar holds when one permits sources, as long as one allows both electric and magnetic sources; in that case, the duality also involves a transformation taking electric sources to magnetic sources, and vice versa.}  It follows that if $F_{ab}$ is a source-free solution to Maxwell's equations, then so is $\star F_{ab}$.  One way to see how this duality acts is to fix an observer and decompose $F_{ab}$ at a point, relative to that observer's state of motion, into electric and magnetic fields $E^a$ and $B^a$.  In that case, the duality acts as $(E^a,B^a)\mapsto (B^a,-E^a)$.

We can express this relationship in the framework of categorical equivalence.  Consider again the category $\mathcal{EM}$ described above. Then electromagnetic duality defines a functor $D:\mathcal{EM}\rightarrow\mathcal{EM}$, as follows: $D$ acts on objects as $F_{ab}\mapsto \star F_{ab}$; and on arrows as $\chi\mapsto \chi$.\footnote{That $D$ is a functor follows from the fact that each map $\chi$ is an isometry, and the action of isometries commutes with the Hodge star operation.  We stress that $D$ is a covariant functor, and thus cannot realize a duality in the category theoretic sense.}  Defined thus, it follows immediately that $D$ is an \emph{autoequivalence} of $\mathcal{EM}$.\footnote{In fact, it is an automorphism---i.e., a categorical isomorphism from $\mathcal{EM}$ to itself.  To see this, just note that every $F_{ab}$ arises as the Hodge dual of some $F_{ab}$---namely, $\star F_{ab}$---and that $D$ is clearly full and faithful because of its trivial action on arrows.}

I can now state my central claim more precisely.  The functor $D$ does not preserve empirical content, in the follow sense.  Recall that in the previous section, I defined an ``empirical significance'' functor, $E:\mathcal{EM}\rightarrow W$, that takes Faraday tensors to stylized phenomena.  The claim is that $D$ does not commute with $E$: $E\circ D\neq E$.  In other words, the electromagnetic duality maps Faraday tensors to Faraday tensors that give rise to \emph{different} predictions.  The theories are, in this sense, empirically distinguishable, and thus cannot be theoretically equivalent.

\section{Manna lelyalde duality?}

Thus far we have argued that dualities do not (necessarily) preserve empirical content, in the following sense: fix, in advance, how one's mathematical formalism(s) will correspond to observable phenomena; then the duality transformation will not necessarily preserve those phenomena.  We will now consider some reactions one might have to this argument.

One reaction is to say that we should not apply the same ``empirical significance'' functor both before and after applying the duality.  Instead, to capture the sense in which the duality realizes an empirical equivalence, one needs to consider both our original empirical significance functor, and another, ``dual'' empirical equivalence functor that properly assigns predictions to Faraday tensors under the duality.

One consideration strongly favoring this perspective is that the stylized predications of electromagnetism, as described above, involve theoretical concepts that should be transformed under the duality.  In particular, the predictions concern the trajectories of (electrically) charged ``test particles''.  And charge-current density should not be invariant under the duality: though we have been considering the vacuum case so far, were we to include sources, one would expect the duality to act by taking electric charge-current density to \emph{magnetic} charge-current density, and vice verse.  Though test matter is not, by definition, source matter, one might nonetheless insist that it is an inconsistent application of the duality transformation to keep our test matter fixed while transforming fields (and, if appropriate, sources).  Thus, the argument goes, what we really need is an empirical significance functor $\tilde{E}$ that associates each Faraday tensor with the trajectories of accelerated \emph{test (magnetic) monopoles} instead of test charges.

This response has merit, but I see at least two counterarguments.  The first is that employing this dual empirical significance functor seems in tension with a natural rough-and-ready criterion of empirical \emph{in}equivalence, where two theories are empirically inequivalent if I can build a device that would permit me to empirically distinguish them.  In the present case, one \emph{could} build such a device: in effect, one could build a box that, on its outside, had a digital display giving us a value of the electric and magnetic fields inside the box.\footnote{The box might work by, for instance, measuring the force acting on bodies with known charge-to-mass ratio.  But the details of its operation are irrelevant to the point, which is that actual measuring instruments may realize $E$ or $\tilde{E}$, but not both.}  And any such device, if used to measure the electric and magnetic fields associated with a Faraday tensor $F_{ab}$ and its dual $\star F_{ab}$ would, in general, yield different results.  In this sense the theory and its dual are empirically distinguishable by direct experimental test.

Second, while it is true that one can always find a different ``empirical significance'' functor, $\tilde{E}$, with the property that $E = \tilde{E}\circ D$.  In fact, that $D$ is an automorphism of the category $\mathcal{EM}$ ensures that this functor $\tilde{E}$ is uniquely given by $\tilde{E} = E\circ D^{-1}$.  If we use this ``dual'' empirical significance functor, we regain a sense in which the duality preserves empirical significance.  But we should be cautious about this conclusion, because, as is evident from the definition of $\tilde{E}$, we get to claim equivalence only by \emph{inverting} the duality and then applying our old empirical equivalence functor, $E$.  In other words, we get equivalence only by systematically re-defining all of the terms so as to effectively ``undo'' the duality.  We can indiscriminately swap electric and magnetic fields as long as we simultaneously redefine ``electric field'' to refer to what we used to mean by ``magnetic field''; redefine ``magnetic field'' to mean what we used to mean by (negative) ``electric field''; ``magnetic monopole'' to what we used to mean by ``point charge''; and so on.\footnote{\citet{Huggett}, too, considers the possibility that dual theories are (empirically) equivalent only insofar as we take the duality mapping to be a ``reinterpretation'' or ``translation manual'' generating a new, dual semantics.  But he ultimately rejects this view, arguing (after \citet{Huggett+Wuthrich} and \citet{Matsubara}; see also \citet{LeBihan+Read}) that there is another way of seeing dual theories as equivalent, namely, by taking only those claims on which dual theories \emph{agree} (on the original semantics) to be physically significant.  On this view, one should think of a duality as a ``symmetry'' and then ``quotient'' out by the duality map.  It is hard to see \emph{any} claims that are preserved by the electromagnetic duality functor, but pursuing this further will take us beyond the scope of this paper.}

Given these objections, the sense of empirical equivalence given by appealing to $\tilde{E}$ might seem trivial.  And in a sense it is: if we permit arbitrary re-interpretations of our theories to \emph{force} empirical equivalence whenever we have some mathematical equivalence, then empirical equivalence adds nothing to a (sufficiently strong) notion of mathematical equivalence.  But this is not to say that the duality is trivial.  In fact, that the duality map exists reflects a highly non-trivial mathematical fact about the formalism of electromagnetism---and, likewise, for the more sophisticated dualities described above.  Even the existence of $\tilde{E}$ is non-trivial, insofar as it captures the sense in which, given the duality $D$, it is possible to systematically and completely perform the re-interpretation necessary to give a different, but, in a sense, equally good, recipe for extracting empirical significance from the formalism of electromagnetism.

There is another line of objection that is related to the foregoing, but which I think is stronger.\footnote{I am grateful to Ben Feintzeig for pressing me on this point (over and over again).}  One might think that there is a basic problem with the starting point for the analysis given here.  Following the empirical equivalence literature, I have assumed that it is possible, given a theory, to fix, once and for all, how that theory makes predictions.  In fact, how we interpret the mathematics we use in physics itself depends on contingent facts about physical systems, relationships between different sets of laws, and in many cases, the history of how we developed the mathematics for particular applications.  It is not something we ``fix'', by fiat, given some collection of mathematical structure; nor is it something we can arbitrarily revise.  Now, after the long and messy work of building theories, applying them, and gradually learning how to interpret them is done, one might be able to summarize the results using something like the functor $E$.  But $E$ is not given to us, prior to and independent of that work.  From this perspective, then, it is wrong-headed to try to use $E$ to extract empirical information from the dual theory, $D(\mathcal{EM})$, because to do so is to stipulate meanings for a putatively ``new theory''; really, one would need to imagine trying to apply the dual theory to real world phenomena, and ask how we would come to understand how to interpret the formalism from that starting point.  Doing this would very likely lead one to $\tilde{E}$ as a natural way of association the formalism with physical situations.\footnote{Why?  At least in our own world $\tilde{E}$ is strongly preferred on the grounds of the non-existence of magnetic monopoles.}  So taking $E$ to reflect the ``default'' interpretation of the formalism is misguided.

This response further supports the suggestion above, that moving to a different interpretation may be well-motivated in some cases.  But it does not undermine the claim that $E$ and $\tilde{E}$ are different---or that we could not empirically distinguish electromagnetism from its dual.

Where does this leave us?  The upshot is that there is a sense in which dual theories are empirically equivalent, and a (stronger) sense in which they are not, at least in general.  They are not (in general) empirically equivalent in the sense usually required for theoretical equivalence: they do not necessarily preserve ``prior'' empirical content.  They are empirically equivalent in a weaker sense, where we permit re-interpretations that force the equivalence.  In other words, a theory is, in general, equivalent to its dual theory only with a compensating ``anti-dual'' empirical interpretation.  How well-motivated this re-interpretation is will depend on context.

\section*{Acknowledgments}
I am grateful to Seb de Haro for organizing the session in which this paper was presented, and to my co-symposiasts, Thomas Barrett, Seb de Haro, and Laurenz Hudetz for a lively and interesting session.  I am also grateful to Sean Carroll, Ben Feintzeig, Nick Huggett, and Sarita Rosenstock for very helpful comments on the talk, and to Seb de Haro for detailed comments on a draft.   This paper was made possible in part through the support of grant \#61048 from the John Templeton Foundation. The opinions expressed in this publication are those of the author and do not necessarily reflect the views of the John Templeton Foundation.

\bibliography{duality}
\bibliographystyle{elsarticle-harv}
\end{document}